\documentstyle[prl,twocolumn,epsfig,psfrag,floats,aps]{revtex}

\newcommand{\be}{\begin{equation}}
\newcommand{\ee}{\end{equation}}
\newcommand{\bea}{\begin{eqnarray}}
\newcommand{\eea}{\end{eqnarray}}

\newcommand{\wtPsi}{\widetilde{\Psi}}

\newcommand{\bS}{{\bf S}}
\newcommand{\br}{{\bf r}}

\newcommand{\nonu}{\nonumber\\}

\begin{document}

\twocolumn[\hsize\textwidth\columnwidth\hsize\csname
@twocolumnfalse\endcsname

\title{Quantum Hall states and boson triplet condensate
       for rotating spin-1 bosons}
\author{J.W.~Reijnders, F.J.M.~van Lankvelt, K.~Schoutens}
\address{Institute for Theoretical Physics, University of Amsterdam,
Valckenierstraat 65, 1018 XE Amsterdam, The Netherlands \\
Department of Physics, P.O.Box 400714, University of Virginia,
Charlottesville, VA 22904-4714, USA}
\author{N.~Read}
\address{Department of Physics, Yale University, P.O.Box 208120,
New Haven, CT 06520-8120, USA}
\date{June 3, 2002}
\maketitle

\begin{abstract}
We propose and analyze two series of clustered quantum Hall states
for rotating systems of spin-1 bosons. The first series (labelled
SU$(4)_k$) includes the exact ground states of a model Hamiltonian
at large angular momentum $L$, and also for $N=3k$ particles at
$L=N$. The latter is a spin-singlet boson triplet condensate
(BTC). The second series, labelled SO$(5)_k$, includes exact
ground states at large $L$ for different parameter values.
\end{abstract}


%
%


\vspace{0.1in} ]

In the study of the phenomenon of Bose-Einstein condensation (BEC)
of cold atoms, the possibility of using bosons with internal
degrees of freedom (spin 1 in particular), and of rotating the
condensate has led to fascinating new possibilities for the states
of matter that can be formed and observed. A multicomponent BEC
can be realized by trapping higher spin atoms such as ${}^{87}$Rb
\cite{myatt} and  ${}^{23}$Na \cite{sta-k} in optical traps, which
do not affect the spin degeneracy, and spin-1 or ``vector'' BEC's
have been realized and studied \cite{sta-k,barrett}. In a model
description, the 2-body interactions of the spin-1 atoms are
written as $\delta(\br_1-\br_2)[c_0+c_2 \bS_1 \cdot \bS_2]$, with
$c_0=(g_0+2g_2)/3$, $c_2=(g_2-g_0)/3$, and $g_S =
4\pi\hbar^2a_S/M_b$ with $M_b$ the boson mass and $a_S$ the
$s$-wave scattering length in the total spin $S$ channel
\cite{om,Ho}. In the case of ${}^{23}$Na one has $c_2>0$, and
hence ``anti-ferromagnetic'' spin correlations. The ``polar''
ground state \cite{om,Ho} has $\langle \bS \rangle=0$; it supports
many interesting collective excitations (see \cite{Zh} for a
review). The case of ${}^{87}$Rb has $c_2<0$, leading to a
spin-polarized (ferromagnetic) ground state.

The effect of rotating a ``scalar'' BEC of spinless or
spin-polarized atoms in an isotropic harmonic trap of frequency
$\omega_0$ can be studied at mean field level (the
Gross-Pitaevskii equation), and a vortex lattice is found at
sufficiently high rotation frequency $\omega$, in agreement with
experimental observations \cite{vortex-exp}. At still higher
frequencies, quantum fluctuations become important and it has been
argued that the vortex lattice is replaced by a sequence of
distinct quantum fluids, which can be understood using a mapping
to the fractional quantum Hall (qH) effect \cite{WGS}. In this
paper, we explore the corresponding possibilities in a rotating
system of spin-1 bosons.

In the regime $\omega_0-\omega\ll \omega_0$ and very weak
interaction, only the lowest-energy single-particle states in the
trap are of interest. These have non-negative angular momentum
$m=0$, $1$, \ldots about the rotation axis, and no excitation of
the motion along the axis, so the motion is effectively in the
$xy$-plane. The $x$, $y$ dependence of these wave functions is
$\psi_m(z) \propto z^m e^{-|z|^2/2}$ (where $z=x+iy$, and the
quantum length $(\hbar/M_b\omega_0)^{1/2}$ of the trap, and
$\hbar$, have been set to 1); they correspond to the lowest Landau
level (LLL) \cite{WGS}.

We consider the interaction Hamiltonian for bosons $i$,
$j=1$, \ldots, $N$, restricted to the LLL, %
\be H_{\rm int}=\sum_{i< j} \delta(\br_i-\br_j)[c_0+c_2 \bS_i
\cdot \bS_j] \ .
\label{eq:int02} \ee %
The ground state (GS) for the system rotating at frequency
$\omega$ is found by minimizing $H_{\rm tot}=H_{\rm int} +
(\omega_0-\omega) L$ \cite{WGS}, where $L=\sum_i m_i$ is the total
angular momentum. For $N$ particles in the LLL with zero
center-of-mass angular momentum, the average ``filling factor'' of
the occupied states can be defined as $\nu=N/(N_V+1)$, where
$N_V=2L/N$ is the number of vortices there would be if there were
a Bose condensate (we assume here a uniform average occupation of
states up to $m = N_V$). The use of only the LLL states is
physically reasonable when $\omega_0-\omega \ll \omega_0 /N_V$ (so
that particles in the LLL with $m=N_V$ cannot lower their energy
by moving to a non-LLL state with $m=0$) and $\nu c_S \ll
\omega_0$, $S=0$, $2$ (so that perturbative corrections from
mixing non-LLL states into the GS are negligible).


For the scalar case (in the LLL, without $c_2$), a study
\cite{CWG} using periodic boundary conditions found that the
vortex lattice melts at a critical $\nu=\nu_c\sim 10$, that
incompressible fluid states occur at $\nu={1 \over 2}$, $1$, ${3
\over 2}$, $ 2$, \ldots $< \nu_c$, and that the corresponding GS
wavefunctions have large overlaps with the Read-Rezayi (RR)
\cite{RR2} series of fractional qH states.

Here, for the case of vector bosons, we first propose two series
of spin-singlet generalizations of the RR states (labelled
SU$(4)_k$ and SO$(5)_k$, respectively, and $k=1$, $2$, \ldots),
and explain their structure. In a thermodynamic limit $N\to
\infty$ with $\nu$ fixed, they each represent incompressible
liquid phases of the system, with $\nu=3k/4$ for the SU$(4)_k$
states and $\nu=k$ for the SO$(5)_k$ series. Then we present
results for the GS's of $H_{\rm int}$ as a function of $L$, $c_0$,
and $c_2$. We identify a region in the $c_0$--$c_2$ plane where
the proposed qH states with $k=1$ give the exact GS's at high
values of $L$. We demonstrate that the $k=N/3$ member of the
SU$(4)_k$ series is an exact eigenstate and, apparently, the GS of
$H_{\rm int}$ with $c_2=0$, $c_0>0$, at angular momentum $L=N$. In
second quantization, this state, a spin-singlet boson-triplet
condensate (BTC), takes the
form %
\be | {\rm BTC} \rangle =
[\epsilon^{\mu_1\mu_2\mu_3} b_{0\mu_1}^\dagger b_{1\mu_2}^\dagger
b_{2\mu_3}^\dagger]^{N/3}|0\rangle \ , \label{eq:btc} \ee %
where $b_{m\mu}^\dagger$ is a creation operator for the
single-particle state $\psi_{m}(z)$ with spin $\mu$, where
$\mu=x$, $y$, $z$. Finally, we show the GS's of $H_{\rm tot}$ as a
function of $\omega$.

We now explain our analysis, starting with the construction of
spin-singlet qH states for spin-1 bosons. This construction is a
generalization of similar constructions for scalar and spin-1/2
particles (either of which could be electrons or bosons)
\cite{RR2,ARRS,ALLS}. These papers all employed a correspondence
\cite{MR} between qH states and specific conformal field theories
(CFT's). This connection allows one to obtain trial qH
wavefunctions as chiral correlators in CFT's that are associated
to a Lie algebra $G$ and integers $k\geq 1$ and $M\geq 0$, where
$M$ is even (odd) for bosons (fermions). These states generalize
the pairing familiar from the theory of superconductivity to the
formation of ``clusters''. The structure of these CFT's is such
that the resulting qH wavefunctions have a property that
guarantees that they are exact zero-energy eigenstates of certain
model Hamiltonians. For example, starting from $G=$ SU$(n+1)$ and
putting $M=0$, one can find completely symmetric wavefunctions
that obey \bea \wtPsi(z_1,z_2,\ldots,z_N) \neq 0 & \hbox{ for } &
z_1=\ldots=z_l \ (l\leq k), \nonu \wtPsi(z_1,z_2,\ldots,z_N) = 0 &
\hbox{ for } & z_1=\ldots=z_{k+1} \ , \label{eq:clus} \eea
independent of the spins of the particles involved. Hence, these
are zero-energy eigenstates of a Hamiltonian with a repulsive
$k+1$-body $\delta$-function interaction \cite{RR2}, and are the
unique states of the lowest $L$ with this property. The existence
of such a Hamiltonian and the fact that it has a gap in its energy
spectrum ensures that a corresponding incompressible liquid phase
of matter exists (when $N\to\infty$ at fixed $k$, $M$) over a
range of interaction parameters, not just for the $k+1$-body
model. The filling factors of these SU$(n+1)_k$ states are \be
\nu(n,k,M) = {nk \over n k M+ n +1} \ . \label{eq:nu-nkm} \ee The
parameter $n$ corresponds to the number of components of an
internal degree of freedom of the particles, and these models have
U$(n)$ symmetry; the clusters contain $nk$ particles. For $n=1$
this construction gives the RR states, while for $n=2$ it produces
spin-singlet states for spin-1/2 particles \cite{ARRS}. Finally,
clustered qH states admit excitations (quasiparticles) of
effectively fractional ``charge'' (i.e.\ particle number). The
simplest of these can be viewed as the result of the adiabatic
insertion of a fraction of a quantum of magnetic flux (or
vorticity), where the allowed fraction is $1/nk$ in the present
cases. States with more than three quasiparticles at
well-separated fixed positions display large degeneracies, which
give rise to ``non-abelian statistics'' \cite{MR} and are
understood for $n=1$, $2$ \cite{ARRS,Ar}.

Our first series of clustered spin-singlet qH states of spin-1
bosons consists of the SU$(4)_k$ states ($n=3$ above), where we
put $M=0$. The filling factor is $\nu=3k/4$. The CFT construction
guarantees that, for $N$ divisible by $3k$, this state is a
singlet under SU(3), and hence also under its SO($3$) subgroup,
the usual spin rotation group. The wavefunction for $N=3kp$
particles can be written in terms of its components for particular
spin states $\mu=x$, $y$, $z$ specified for each particle (similar
to \cite{CGT-ALS}), as %
 \bea && \wtPsi^{\rm SU(4)}_k
   = {\cal S}_{\rm groups} \left[ P_{\rm groups}
   \wtPsi^{2,2,2,1,1,1} \right],
\label{eq:su4-k} \eea where %
\bea \lefteqn{ \wtPsi^{2,2,2,1,1,1}
(z^x_1,\ldots,z^x_p;z^y_1,\ldots,z^y_p;z^z_1,\ldots,z^z_p) =}
\nonu && \quad \prod_{\mu=x,y,z} \prod_{i<j}(z^\mu_i-z^\mu_j)^{2}
\prod_{\mu'<\mu''} \prod_{i,j}(z^{\mu'}_i-z^{\mu''}_j) \ .
\label{eq:222111} \eea In words, the operations $P_{\rm groups}$
and ${\cal S}_{\rm groups}$ in eq.\ (\ref{eq:su4-k}) tell one to
divide the $3pk$ bosons into $k$ groups of $3p$ bosons each ($p$
of each spin polarization), to write a factor
$\wtPsi^{2,2,2,1,1,1}$ as in eq.\ (\ref{eq:222111}) for each
group, and finally, to symmetrize over all ways the particles can
be divided over the $k$ groups. For $k=1$ there is a single group,
and one finds a state of total degree $L=3p(2p-1)$ that
generalizes the Laughlin and Halperin states for scalar and
spin-$1/2$ particles, respectively, and which is a zero-energy
eigenstate of $H_{\rm int}$ for all $c_0$, $c_2$, and the unique
GS when both $g_0$, $g_2 > 0$ since $H_{\rm int}$ is then
positive. Putting instead $k=N/3$ gives $k$ groups of $3$
particles each, the resulting state being the $L=N$ BTC, eq.\
(\ref{eq:btc}). The fundamental quasiparticles over the SU$(4)_k$
qH liquids have fractional ``charge'' equal to $\pm 1/4$, and spin
1, for all $k$. We note however that such assignments, while
meaningful for $N\gg k$, may be meaningless when $N\sim 3k$, where
the quasiparticle size is comparable to the size of the fluid
drop.

An important point is that for $c_2=0$, the Hamiltonian
(\ref{eq:int02}) has SU$(3)$, not just SO$(3)$, spin-rotation
symmetry. Therefore, we expect the SU$(4)_k$ states to be relevant
near this line when $c_0>0$, for sufficiently large $L$.

Our second series of qH states for spin-1 bosons is obtained from
a CFT with $G=$ SO$(5)$, and generalizes the construction in Ref.\
\cite{ALLS}. The GS's possess a symmetry under an SO$(3)$ subgroup
of SO$(5)$. The general construction, which involves spin-singlet
clusters of $2k$ particles, gives states of filling factor
$\nu=k/(kM+1)$. Putting $M=0$, we obtain qH states with $\nu=k$.
For $k=1$, the wavefunction with values in the spin space of $N$
spin-1 particles can be written as \be \wtPsi^{\rm
SO(5)}_{k=1}(z_i) = {\rm Pf} \left[
{\rho_i\rho_j+\sigma_i\sigma_j+\tau_i\tau_j \over (z_i-z_j)}
\right] \wtPsi^1_{\rm L}(z_i) \ . \label{eq:so5-k}
\ee %
Here $\wtPsi^1_{\rm L}(z_i)=\prod_{i<j}(z_i-z_j)$ is a
spin-independent Laughlin factor in all $N$ particle coordinates;
the Pfaffian $\rm Pf$ is defined, as usual, by
\begin{equation}
{\rm Pf}\,\left(M_{ij}\right)={\cal A}\left( M_{12}M_{34}\ldots
M_{N-1,N}\right),
\end{equation}
where $M_{ij}$ are the elements of an antisymmetric matrix, and
$\cal A$ denotes the operation of antisymmetrization; $\rho_i$,
$\sigma_i$, $\tau_i$ are basis vectors in the spin space for
particle $i$, that correspond to $x$, $y$, $z$, and the product is
the tensor product.  For $k=1$, $N$ must be even, and the
clustering (or pairing) of particles is seen explicitly. Clearly,
there must be even (but otherwise arbitrary) numbers of particles
in each of the three spin states $\rho$, $\sigma$, $\tau$.  The
$k=1$ wavefunction for $N=2q$ particles has total degree
$L=2q(q-1)$. The function (\ref{eq:so5-k}) can be viewed as
spin-singlet p-wave pairing of composite fermions of spin 1 (for a
review, see Ref.\ \cite{read01}). There are also excited states
with unpaired ``charge''-neutral spin-1 fermions. We note that, in
the state (\ref{eq:so5-k}), two bosons are found at the same point
only if the total spin of the pair is zero, not if it is 2, and
hence the state is an exact zero-energy eigenstate, of lowest $L$,
for a $\delta$-function interaction that includes a projection
onto spin 2. In our parameterization, that is $c_2=c_0/2$ or
$g_0=0$, and if also $c_0>0$, then $H_{\rm int}$ is positive, so
our $k=1$ state is the GS at this $L$. For general $k$, the
SO$(5)_k$ wavefunction can be written as a CFT chiral correlator,
and is an exact zero-energy eigenstate of a certain $k+1$-body
$\delta$-function interaction (details will appear elsewhere). In
general, the quasiparticles over the SO$(5)_k$ qH state have
``charge'' $\pm 1/2$ and spin $1/2$, thus displaying a
fractionalization of both ``charge'' and spin.

\begin{figure}
\psfrag{g1}{$g_0=0$}
\psfrag{g2}{$g_2=0$}
\psfrag{S3}{$SU(3)$}
\psfrag{c2}{$c_2$}
\psfrag{c0}{$c_0$}
\psfrag{II}{II}
\psfrag{Ia}{Ia}
\psfrag{Ib}{Ib}
\begin{picture}(75,150)(-20,-5)
\epsfig{file=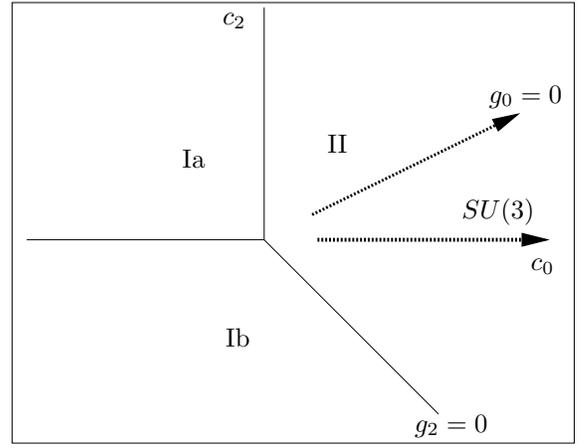,width=75mm}
\end{picture}
\caption{Overview of $c_0$--$c_2$ plane, with
special regions and directions marked.}
\label{fig:c-plane}
\end{figure}

We next report on our study (numerical and analytic) of the GS
phase diagram of the model $H_{\rm int}$. The numerical work is
restricted to small values of $N$ (up to 12), but it indicates
many features which we believe to hold for general values of $N$.
To guide our discussion we have displayed in
Fig.~\ref{fig:c-plane} various special directions (rays) and
regions in the $c_0$--$c_2$ plane.

For $L=0$, the GS has total spin $S=N$ for $c_2<0$ (ferro regime)
and $S=0$ ($1$) for $N$ even (resp.,\ odd) for $c_2>0$ (anti-ferro
regime). For $c_2=0$, there is a single SU$(3)$ multiplet of spin
states, decomposing into unique SO($3$) multiplets of each spin
$S=N$, $N-2$, \ldots.

As $L$ increases, these two phases survive in part of the phase
diagram, as compact drops of fluid, with the center of mass
carrying all the angular momentum. Meanwhile, the positive $c_0$
axis gradually opens into a region that contains other phases. By
the time $L$ is $\geq N$, the $c_0$--$c_2$ plane contains the
three regions labelled Ia, Ib, and II in Fig.~\ref{fig:c-plane}.
The GS's in regions Ia and Ib are similar to the GS in the
``attractive'' regime in the scalar case \cite{WGS}. The orbital
part of the GS wavefunction is of the form $\wtPsi(z_i) \propto
z_c^L$, with $z_c=\sum_i z_i/N$ the center of mass (CM)
coordinate. In region Ia ($c_0<0$, $c_2>0$), the spin state is the
same spin-singlet as for the $L=0$ GS, and the GS energy becomes
$[c_0 N(N-1)/2 - N c_2](2\pi)^{-3/2} $ \cite{lawho}. In region Ib
($c_2<0$, $c_0<-c_2$), the spin state is ferro, $S=N$, giving
energy $(c_0+c_2)N(N-1)(2\pi)^{-3/2}/2 < 0$ \cite{WGS}. At
$c_2=0$, $c_0<0$, the spin states again form the SU$(3)$
multiplet. In the ``repulsive'' region II, the GS is in general
not a common eigenstate of the $c_0$ and $c_2$ parts of the
interaction, and the GS energy depends non-linearly on the ratio
$c_2/c_0$. The $L=N$ GS's all have $S=0$ or $S=1$ (depending on
the value of $N$ modulo 6, and on the ratio $c_2/c_0$). For $L>N$,
larger values of $S<N$ do occur near the boundary at $c_2=-c_0$
($g_2=0$).

Turning to $L\gg N$ in region II, we have already pointed out that
the SU$(4)_1$ state is the zero-energy GS for $N=3p$ particles at
$L=3p(2p-1)$ when $g_0$ and $g_2 > 0$, while the SO$(5)_1$ state
is the zero-energy GS for $N=2q$ particles at $L=2q(q-1)$ for
$g_0=0$. (For $N=3$, $4$, these states occur at $L=N$). For even
larger $L$, each of these model cases possesses many degenerate
GS's of zero energy. This implies that, within our model, the
SU$(4)_1$ and SO$(5)_1$ GS's are those found (for the parameters
as stated) at the critical rotation frequency $\omega=\omega_0$,
and so the lowest possible filling factor is $3/4$ in the region
$g_0$, $g_2 >0$, but is $1$ when $g_0=0$.

At intermediate $L>N$ values in this region, we expect similar
physics. Thus, for $c_2=c_0/2>0$, ($g_0=0$), the system can lower
its energy by forming SO$(3)$ singlet pairs of bosons. For
$L<2q(q-1)$ ($N=2q>4$) we do not have exact eigenstates, but we
expect that, similar to Ref.\ \cite{CWG}, for $\nu=k$ less than
some critical value $\nu'_c>1$, the bulk of the fluid will be in
the SO$(5)_k$ state. For $c_2=0$, $c_0>0$, the preferred behavior
is singlet formation via triples of bosons of spin zero; each such
three bosons must be in an antisymmetric orbital state as well.
For $L<3p(2p-1)$ ($N=3p>3$), we do not generally have exact
eigenstates, but we expect the bulk of the fluid to be the
SU$(4)_k$ states for $\nu=3k/4$ less than another critical
$\nu_c''>3/4$.

We also found that the exact GS for $N=6$, $L=6$ at $c_2=0$ is the
SU$(4)_2$ or BTC state (\ref{eq:btc}). Prompted by this, we have
proved analytically that for $N=3p$ particles the SU$(4)_p$ state
is an exact eigenstate of $H_{\rm int}$ with $c_2=0$, with
eigenvalue \be E = {11 \over 48}N(N-3)(2\pi)^{-3/2}c_0 \ .
\label{eq:EBTC} \ee We have numerically checked that for $N=3$,
$6$, $9$, $12$ this state is the GS and believe this is true for
all $N=3p$. [The energy eq.~(\ref{eq:EBTC}) is lower than the
value ${1 \over 4} N(N-2)(2\pi)^{-3/2}c_0$ of an SU$(3)$ multiplet
of vortex states for spin-1 bosons, that have the same orbital
wavefunction as the $c_0>0$ vortex state in the scalar case.]

\begin{figure}
\psfrag{n}{$\langle n\rangle$}
\psfrag{x}{$x$}
\begin{picture}(60,100)(-30,-8)
\epsfig{file=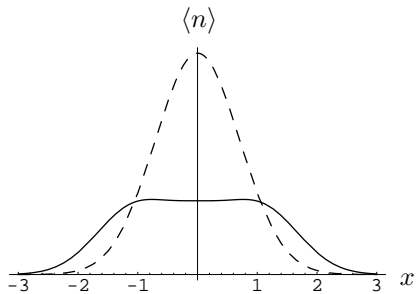,width=60mm}
\end{picture}
\caption{Density profiles versus $x$ at $y=0$ in the LLL model:
BTC (solid line); non-rotating BEC (broken line). Vertical axis is
in units of the particle number $N=3p$.} \label{fig:density}
\end{figure}

The BTC state in eq.\ (\ref{eq:btc}) has a clear experimental
signature, that is the density profile given in
Fig.~\ref{fig:density}, whose form is independent of $N$. We
caution again that one should probably not think of the limit
$N=3k$ as a qH state, since the drop is so small. We identified
the state as an extreme member of the SU$(4)_k$ qH series, but it
might be more useful to view it as a boson-triplet analog of the
BCS paired electron states, or alternatively as an SU$(3)$ analog
of a skyrmion spin texture (see Ref.\ \cite{mmk}).

\begin{figure}
\psfrag{L}{$L$} \psfrag{w}{$\omega$} \psfrag{N}{$N=6$}
\psfrag{c}{$c_0>0,\,c_2=0$}
\begin{picture}(80,110)(0,-5)
\epsfig{file=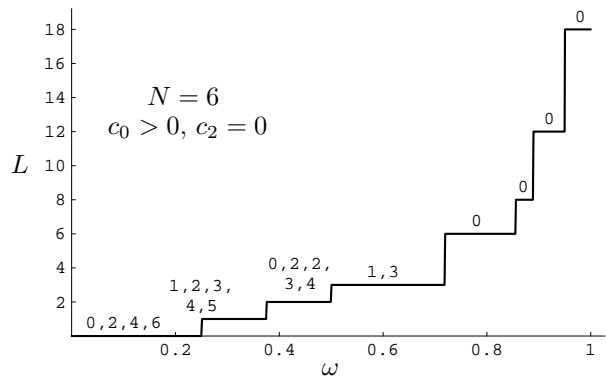,width=80mm}
\end{picture}
\caption{GS values for $L$ as a function of the rotation frequency
$\omega$ for $N=6$ particles at $c_2=0$. (We put $\omega_0=1$ and
$c_0=(2\pi)^{3/2}/4$.) The state at $L=6$ is the BTC, while the
$L=18$ state is the SU$(4)_1$ state. The spin values $S$ are shown
for each GS.} \label{fig:LvsOmega}
\end{figure}

The response of the spin-1 boson system to a rotation frequency
$\omega$ is found by minimizing $H_{\rm tot}$. In
Fig.~\ref{fig:LvsOmega} we display the result for $N=6$ particles.
The BTC is at $L=6$, and the SU$(4)_1$ state is at $L=18$. The
degenerate spin multiplets listed at the steps at $L<6$ each form
a single SU(3) multiplet. The wavefunctions of these GS's at
$L\leq N$ are uniquely determined by their SU$(3)$ and $L$ quantum
numbers.

The SO$(5)_1$ state found here will survive for some distance off
$g_0=0$. For $N$ large at fixed $\nu$, there will be several
phases within region II, and in particular a boundary between the
SO$(5)_1$ and SU$(4)_1$ phases that approaches $g_0=0$ as $\omega
\to \omega_0$.

Our constructions for $c_2=0$ directly generalize to an
$n$-component rotating Bose gas with repulsive spin-independent
$\delta$-function interactions, implying SU$(n)$ symmetry. In
particular, we expect a ``boson $n$-plet condensate'' at
$L=(n-1)N/2$, and SU$(n+1)_k$ states at sufficiently small
$\nu=nk/(n+1)$.

To conclude, we have found several interesting states of matter,
not containing vortices, in rotating spin-1 bosons.

As this manuscript was prepared, two preprints appeared
\cite{pzc,hm} that also address a rotating BEC.
%
%
These papers point out the relevance of the SU$(n+1)_1$ states at
large $L$, but do not discuss the other qH states or the BTC state
that we consider.


We thank E.~Rezayi, M.~Kasevich, S.M.~Girvin, T.-L.~Ho, and
E.~Mueller for helpful discussions. This research was supported by
the Netherlands Organisation for Scientific Research, NWO, and the
Foundation FOM of the Netherlands (JWR, FJMvL and KS), and by the
NSF under grants nos.\ DMR-98-18259 (NR) and DMR-98-02813 (KS).

\end{document}